\RequirePackage[mathlines]{lineno}
\documentclass[12pt]{iopart}

\usepackage{graphicx}
\usepackage{amssymb}
\usepackage{mathrsfs}
\usepackage{verbatim}
\usepackage{color}
\usepackage{algorithm,algorithmic}
\usepackage{cite}
\usepackage{url}
\usepackage{harvard}
\usepackage{iopams}
\citationmode{abbr}
\citationstyle{dcu}

\def\cT{{\mathcal T}}
\def\R{{{\mathbb R}}}

\begin{document}

\title[Optimal Marker Locations for Tumor Motion Estimation]{Optimal Surface Marker Locations for Tumor Motion Estimation in Lung Cancer Radiotherapy}

\author{Bin Dong}
\address{Department of Mathematics, The University of Arizona, Tucson,
AZ, 85721-0089, USA, and Center for Advanced Radiotherapy
Technologies, University of California San Diego, La Jolla, CA 92037-0843, USA}
\ead{dongbin@math.arizona.edu}
\author{Yan Jiang Graves, Xun Jia and Steve B. Jiang}
\address{Center for Advanced Radiotherapy
Technologies and Department of Radiation Medicine and Applied Sciences, University of California
San Diego, La Jolla, CA 92037-0843, USA}

\begin{abstract}	
Using fiducial markers on patient's body surface to predict the tumor
location is a widely used approach in lung cancer radiotherapy. The purpose of this work is to propose an algorithm that automatically identifies a sparse set of locations on the
patient's surface with the optimal prediction power for the tumor motion. In our algorithm, it is assumed that there is a linear relationship
between the surface marker motion and the tumor motion. The sparse
selection of markers on the external surface and the linear relationship between the
marker motion and the internal tumor motion are represented by a prediction matrix. Such
a matrix is determined by solving an optimization problem,
where the objective function contains a sparsity term that penalizes the
number of markers chosen on the patient's surface. Bregman iteration is used
to solve the proposed optimization problem. The performance of our
algorithm has been tested on realistic clinical data of four lung cancer patients.
Thoracic 4DCT scans with 10 phases
are used for the study. On a reference phase, a grid of points are casted on the patient's surface (except
for patient's back) and propagated to other phases via deformable image registration of the
corresponding CT images. Tumor locations at each phase are also manually delineated.
We use 9 out of 10 phases of the 4DCT images to identify a small group of surface
markers that are most correlated with the motion of the tumor, and find the
prediction matrix at the same time. The 10th phase is then used to test the
accuracy of the prediction.
It is found that on average 6 to 7 surface markers are necessary to predict tumor
locations with a 3D error of about 1mm. It is also found that the selected marker locations lie closely in those areas where surface point motion has a  large amplitude and a high
correlation  with the tumor motion.
Our method can automatically select sparse
locations on patient's external surface and estimate a correlation matrix based on 4DCT, so that the selected surface locations can be used to place fiducial markers to optimally
predict internal tumor motions.
\end{abstract}


\submitto{\PMB}

\maketitle

\linenumbers\modulolinenumbers[5]

\section{Introduction}

Modern radiotherapy techniques, such as Intensity Modulated Radiation Therapy
(IMRT), are capable of delivering highly conformal radiation dose to a cancerous
target while sparing critical structures and normal tissues. Intra-fraction
tumor motion caused by patient respiration, however, may lead to geometric
miss of the target and hence potentially compromise the efficacy of these
techniques while treating tumors at lung or upper abdomen
area. To mitigate this problem, a
number of  techniques have been developed, such as gated treatment, for which
accurate modeling and prompt prediction of tumor motion are
necessary \cite{jiang2006technical,Jiang:2006:SRO}.

Tumor localization methods can be generally categorized according to the
locations of surrogates. Methods using internal surrogates, such as gold
markers implanted in or near tumor, are accurate but have issues like the
risks of pneumothorax for lung cancer patients
\cite{Arslan:2002,Geraghty:2003}, marker migration \cite{Nelson:2007}, and
the extra imaging radiation dose \cite{jiang2006technical}. In contrast,
external surrogate based tumor localization is usually noninvasive and
radiation free. In such methods, a regression model is first built between
the coordinates of some empirically selected external surrogates and those of
the tumor using a training data set. Such a model will be utilized  to
predict the tumor location using the real-time measurements of the marker
locations during a treatment via, for example, Cyberknife Synchrony system
(Accuray Corporate, Sunnyvale, CA, USA) \cite{pepin2011correlation}. Yet, the
accuracy of this method usually relies on the correlation between external
marker motion and internal tumor motion for a particular patient
\cite{Hoisak:2004}.

In fact, there are a few questions one should keep in mind while using external markers for tumor tracking. First of all, how many external markers are
necessary? While using more markers may potentially provide more comprehensive information for tumor location estimation, it is evident that the motion of points on a patient surface
is strongly
correlated and information from many surface markers is likely to be redundant.
Clinically, it is necessary and desirable to use a
minimum number of markers to predict the tumor motion to a satisfactory
degree. Second, given the number of markers, where shall we optimally place
them? Despite a lot of studies regarding the patient breathing pattern and the selection of marker locations\cite{Yan:2006,Wu:2008}, markers are placed empirically in most clinical
practice.

In this study, we will attempt to answer the aforementioned two questions
utilizing a sparse optimization approach.  Specifically, our objective is to
choose a sparse set of points from all the points on the front surface of a patient, so that a
linear motion model yields the smallest error in tumor location prediction. With a novel optimization
model to formulate this objective in a clean and precise mathematical
language, as well as an effective numerical algorithm to solve the problem, we can effectively yet
efficiently identify the key surface points used to predict tumor motion. A
linear regression model is also developed during the optimization process,
such that those markers collaboratively predict tumor locations to a
satisfactory extent.

\section{Methods and Materials}

We start with an introduction of some notations. Denote $Y\in\R^{3\times m}$
as a $3\times m$ matrix whose column vectors are the three Cartesian coordinates of the
center of the tumor at various times $t_j$ with $j=1,2,\ldots,m$. Suppose there are $k$ candidate surface points available for tumor motion prediction. We
denote the coordinates of the collection of all of those surface points at
a given time $t_j$ as a column vector $X_j=[\vec x_1(t_j), \vec x_2(t_j),\ldots,\vec x_k(t_j)]^T$, where each vector $\vec x_i = [\vec x_{i1}, \vec x_{i2}, \vec x_{i3}]$ for
$i=1,...,k$ contains three entries corresponding to the three Cartesian coordinates of the point $i$. If we assemble all the
collections of markers $X_j$ associated with different time
$t_j$, we will have the following matrix
$X:=\left[ X_1, X_2,\cdots, X_m\right]\in\R^{3k\times m}$.

\subsection{Optimization Model}
Assume, for simplicity, there is a linear motion model that relates the external marker motion and the tumor motion. Mathematically speaking,  there exist a matrix
$A\in\R^{3\times3k}$ such that $AX\sim Y$. Note that the columns of the matrix $A$ can be also associated to those $k$ surface points, each with three coordinates. If one column of
the matrix $A$ is non-zero, the corresponding coordinate for that surface point is then utilized to predict the tumor motion. As it is our purpose to select only a few surface points
for tumor motion prediction, the problem can be casted as finding a matrix $A$ with only a few non-vanishing columns, such that the motion of tumor recorded in $Y$ can be accurately
characterized by $AX$. Although this is simply a linear motion prediction model, our numerical experiments indicate that such an assumption is reasonable
and leads to accurate tumor location estimations. We shall refer to the
problem of \textit{optimal marker selection as} the problem of finding the
\textit{linear dependence} of the motion of the internal tumor with the
motion of some \textit{sparsely} selected markers.

We propose our \textit{optimal marker selection model} as follows:
\begin{equation}\label{Model:L21Norm}
\min\limits_A\left\{\|A\|_{2,1}:\ AX=Y\right\},
\end{equation}
where $\|A\|_{2,1}:=\sum_j\left(\sum_i a_{i,j}^2 \right)^{\frac{1}{2}}$ and
$A=(a_{i,j})$. In this optimization problem, the objective function is
defined in such a way that it groups all the matrix elements in a column of
$A$ utilizing an $\ell_2$-norm and then takes $\ell_1$-norm among all
columns. Minimizing such an objective function term enables us to enforce
sparsity at only the level of matrix columns. This idea is inspired by that
of compressed sensing \cite{CRT,CT1,CT,Do2}, which is a recent revolutionary
concept in information theory. The applications of such a $\ell_{2,1}$ norm
has been recently explored in many problems, such as beam orientation
optimization for IMRT\cite{Jia:2011:BOO}, to effectively select only a few
groups of elements. Similar idea was also used in \cite{esser2011convex}
where the $\ell_{1,\infty}$ norm was used for matrix factorization with
applications in hyperspectral image unmixing. We remark that the model
(\ref{Model:L21Norm}) not only sparsely selects markers needed to track the
motion of an internal tumor, but also provides the linear dependence of the
motion of the selected markers with that of the tumor at the same time. All
such information is integrated within the solution matrix $A$.


\subsection{Fast Numerical Algorithm}
To solve the proposed optimization problem (\ref{Model:L21Norm}), we use a
Bregman distance-based algorithm proposed by Yin \emph{et. al.} \cite{Yin2008}, which is proven
to be efficient for $\ell_1$ minimization problems. Given matrices $X$ and
$Y$, the fast algorithm that solves (\ref{Model:L21Norm}) can be written into an iterative form as:
\begin{equation}\label{Algorithm:SubMinimization}
\begin{array}{ll}
A^{k+1}&=\arg\min\limits_A\left\{ \mu\|A\|_{2,1}+\frac{1}{2}\|AX-Y^k\|_F^2 \right\},\\
Y^{k+1}&=Y^k+Y-A^{k+1}X,
\end{array}
\end{equation}
where $k$ is the iteration index and $\|\cdot\|_F$ is the Frobenius norm. The optimization problem in the first subproblem of (\ref{Algorithm:SubMinimization}) can be solved
using the proximal forward-backward splitting algorithm
\cite{combettes2006signal,HYZ}, which by itself is an iterative algorithm as:
\begin{equation}\label{Algorithm:FBS}
A^{p+1}=\cT_{\mu}(A^p-\delta(A^pX-Y^k)X^T),
\end{equation}
where $p$ is the iteration index in this subproblem and $\cT_\mu(B)$, for a given matrix $B=[B_1, B_2, \ldots, B_m]$, is
defined as
$$\cT_\mu(B):=\left[\max(|B_1|-\mu,0)\frac{B_1}{|B_1|},\cdots,\max(|B_m|-\mu,0)\frac{B_m}{|B_m|}\right].$$
We note that \cite{Do1,wang2007fast} $\cT_\mu(B)$ is the closed form solution
to $\min\limits_X\left\{\mu\|X\|_{2,1}+\frac{1}{2}\|X-B\|_F^2\right\}.$ For
computation efficiency, we shall not solve the subproblem
(\ref{Algorithm:SubMinimization}) accurately by using numerous iterations of
(\ref{Algorithm:FBS}), but only use one iteration instead. Now, applying
(\ref{Algorithm:FBS}) (with only one iteration) to
(\ref{Algorithm:SubMinimization}), we have the following fast algorithm that
solves (\ref{Model:L21Norm}) (also known as the Bregmanized operator
splitting algorithm \cite{zhang2010bregmanized}):

\begin{algorithm}
\caption{Optimal Marker Selection Algorithm \label{Algorithm}}
\begin{algorithmic}
\STATE{\textbf{Step 0.}} Initialization: $k=0$, $A^0=0$ and $Y^0=0$.
\medskip
\WHILE{stopping criteria is not satisfied}

\STATE{\textbf{Step 1.}} $$A^{k+1}=\cT_{\mu}(A^k-\delta(A^pX-Y^k)X^T)$$

\STATE{\textbf{Step 2.}} $$Y^{k+1}=Y^k+Y-A^{k+1}X$$

\medskip
\ENDWHILE
\end{algorithmic}
\end{algorithm}
The proof of the mathematical properties of this algorithm, such as convergence,  is beyond the scope of this paper. Interested readers can consult references for more
details\cite{Yin2008,zhang2010bregmanized}.

For realistic patient data, because of the presence of noise and the fact that the motion
of internal tumor is only approximately linearly dependent on the external
markers, we should not expect the relative residual $\|A^k X-Y\|_F/\|Y\|_F$
decrease to $0$. In fact, numerically we observe that the relative residual
should have a lower bound whose value depends on $X$ and $Y$ and it is very
difficult to estimate beforehand. Therefore, we adopt the following stopping
criteria: $$\frac{\|A^k X-Y\|_F}{\|Y\|_F}<\epsilon_1\quad\mbox{or}\quad
\frac{\|A^{k-1}-A^k\|_F}{\|A^k\|_F}<\epsilon_2.$$ In other words, we fix an
$\epsilon_1$ as a satisfactory amount for the residual; meanwhile, if such residual
is not attainable, we will terminate the algorithm when $A^k$ is not changing
too much according to the tolerance $\epsilon_2$.

\subsection{Patient Data}
To validate our algorithm with realistic clinical cases,  4DCT scan data of four lung cancer patients is used. For those patients, a four-slice GE LightSpeed CT scanner (GE Medical
Systems, Milwaukee, WI, USA) was used to acquire the 4DCT data for treatment simulation. Each axial CT slice has a thickness of $2.5 mm$ and the 4DCT was obtained using respiratory
signals from the Varian RPM system (Varian Medical Systems, Inc., Palo Alto, CA, USA). The 4DCT scan consists of ten different phases of one breathing cycle; and the CT volume at
each
respiratory phase consists of 100 to 144 slices of CT images covering the most of thorax area depending on patients. Each slice of CT image has $512\times 512$ pixels, with a pixel
size
of $0.977\times 0.977 mm^2$. For each patient, tumor GTV was manually contoured on 4DCT scan images of ten respiratory phases by an expert observer and the 3D tumor center
coordinates
were identified. Table 1 summarizes the number of CT image slices for each CT image volume and the average tumor motion amplitude in the superior-inferior(S-I) direction and average
surface
motion amplitude for each patient. It can be observed that the average tumor motion amplitude in the S-I direction range from $3.3mm$ to $9.0mm$. The average surface point motion
amplitude ranges among all the
patients are found to be $0.8mm$ to $2.0mm$.

\begin{table}[htp]
\centering
\begin{tabular}{c||c|p{4cm}|p{4cm}}
\hline
Patient & No. of slices & Tumor motion amplitude in S-I (mm) & Average surface motion amplitude (mm)\\
\hline\hline
1      & 144  & 6.3  & 2.0 \\
2      & 100  & 9.0  & 1.5 \\
3      & 132  & 7.4  & 1.8 \\
4      & 104  & 3.3  & 0.8 \\
\hline
\end{tabular}
\medskip
\caption{Summary of patient data with number of CT slices, average tumor motion amplitude in S-I direction, and average surface motion amplitude for each patient.}\label{Table:Patient
Data:summary}
\end{table}

Meanwhile, the external surfaces of each patient, excluding the patient's
back, at each phase are extracted by segmenting the CT images using a simple
threshold method. For each patient, the CT image volume at the end of inhale
is set as the target image; the other nine CT image volumes, corresponding to
the other nine different respiratory phases, are set as moving images. The
correspondence between surfaces at different phases is established by
deformable image registration \cite{thirion1998image,Gu2010:PMB:Demons}. When
surface points are available on the external surfaces of each patient, we further
sub-sampled the point sets uniformly to reduce the total number of candidate points for a better computational efficiency. In our
experiments, we choose approximately 200 candidate surface points for each
patient.

\subsection{Validation}
To validate our marker selection algorithm, we
employ an leave-one-out cross validation (LOOCV) method. Specifically, 10 tests are performed for a patient, and for each test, we single out one of the 10 respiratory phases and use
the other 9 to form the matrix $Y$ and solve for the matrix $A$ using
Algorithm \ref{Algorithm}. We then validate our method by using the matrix
$A$ to predict the location of the tumor at the phase that has been singled
out. The deviation of the predicted tumor location from the actual tumor location is characterized by the 3D Euclidean distance between them in mm.

The patients' 4DCT image volumes cover a complete breathing cycle, hence contain information of external surface motion. We could in principle identify regions of interest (ROIs) on
the patient surface that strongly correlate with tumor motions. It is expected that the marker locations selected by Algorithm \ref{Algorithm} should fall closely into those ROIs.
This also serves as a criterion for the justification of the correctness of our marker selection algorithm. To select the ROI, we consider the following two metrics.  First, from
the deformation vector fields between different respiratory phases, the motion trajectory for all surface points were extracted. The correlation function between the internal tumor
motion in the S-I direction and the motion vector of each point on the external surface was employed as a metric. However, only part of the external surface has considerable motion
amplitude and those points with small motion amplitude should not be considered for predicting tumor motions despite their possible high correlations with tumor S-I motion.
Therefore,
we only focus on the surface region with large motion amplitudes. Combining the two criteria, we define the ROI as the areas on the surface in which the motion amplitude is larger than
$80\%$ of the maximum value and the correlation is above 0.85. Although those threshold values for the two criteria are chosen empirically, the general conclusions presented in
the
rest of this paper are found not sensitive to them.

\section{Results}

\subsection{Marker selection}


\begin{figure}[htp]
\centering
    \includegraphics[height=1.5in]{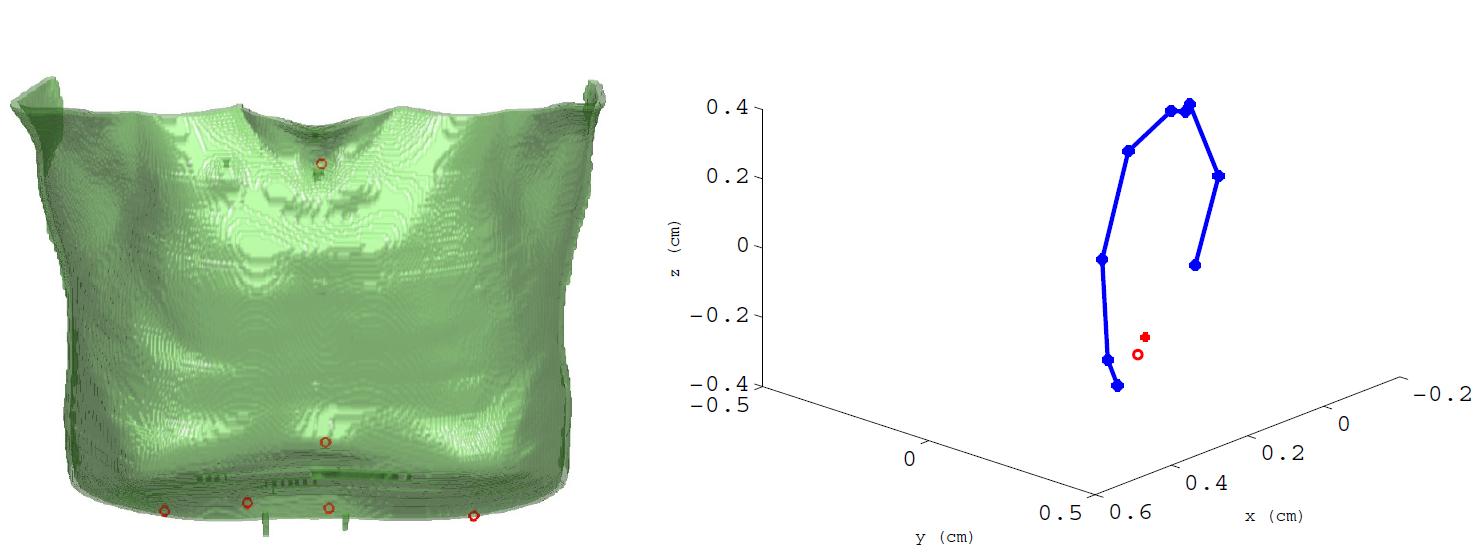}
    \caption{\textit{Left:} Markers
    selected by our algorithm are shown as red circles on one of the patient's
    surface. \textit{Right}: the LOOCV results for the same patient using the phases 1 through 9 as training data (blue dots) and the phase 10 as the testing data (red dot). The red
    circle is the predicted tumor location.
    }\label{Fig:Validation:Real}
\end{figure}

We have studied the validation of our surface marker selection algorithm on 4 lung cancer patients. The selected 6 surface markers in one typical patient (patient No. 4) are drawn in
3D space on the patient surface, as shown in the left panel of Fig.~\ref{Fig:Validation:Real}. Meanwhile, in the right panel of Fig.~\ref{Fig:Validation:Real}, we demonstrate the
LOOCV results for the same patient using the phases 1 through 9 as training data and the phase 10 as the testing data.  Specifically, the blue dots
are the locations of the tumor in the training phases and the red dot is the location of the tumor at the phase 10. The red circle is the predicted location using the selected
surface
markers and the matrix $A$. The 3D distance between the true tumor location and the predicted location is $0.83mm$, indicating the great capability for tumor motion prediction of our
algorithm.

\begin{table}[htp]
\centering
\begin{tabular}{c||c|c|c|c|c|c}
\hline
\multicolumn{1}{c||}{} &
\multicolumn{2}{c|}{Error (mm)} & \multicolumn{2}{c|}{\#Markers} & \multicolumn{2}{c}{Time (sec)}
\\\cline{2-7}
\multicolumn{1}{c||}{Patient} &
\multicolumn{1}{c|}{mean} & \multicolumn{1}{c|}{std} & \multicolumn{1}{c|}{mean} & \multicolumn{1}{c|}{std}
& \multicolumn{1}{c|}{mean} & \multicolumn{1}{c}{std}\\
\hline\hline
 \multicolumn{1}{c||}{1}       & 1.85  & 1.15  & 5.5  & 0.85  & 10.6  &  4.5 \\
 \multicolumn{1}{c||}{2}       & 1.22  & 1.06  & 5.5  & 1.58  &  4.6  &  1.9 \\
 \multicolumn{1}{c||}{3}       & 0.44  & 0.28  & 5.4  & 1.84  & 10.8  &  3.0 \\
 \multicolumn{1}{c||}{4}       & 0.83  & 0.29  & 7.5  & 1.35  & 30.5  & 11.6 \\
 \hline
 \multicolumn{1}{c||}{Average} & 1.08  & 0.69  & 5.98  & 1.04  & 14.1  &  5.2 \\
\hline
\end{tabular}
\medskip
\caption{Summary of tumor location prediction errors, the numbers of
markers selected, and the computation time.}\label{Table:Validation:Real}
\end{table}

A summary of the results of all 10 tests for each of the 4
patients is given in Table~\ref{Table:Validation:Real}. For each patient, we compute the mean and the standard deviation of the 3D errors for the predicted tumor locations and the
number of selected markers over all the 10 tests in the LOOCV. It is found that, on average, our algorithm can automatically select about 6 surface markers that collaboratively
predict tumor motion with an 3D error about $1 mm$.

Algorithm  \ref{Algorithm} is implemented using MATLAB on a laptop with Intel Core i7 (1.73 GHz) CPU and 8.0G
RAM. As for the computation time, it is found that the average time required to perform one optimization is about $14 sec$. We emphasize that the time reported here is
the one for marker selection. Once the markers are selected, the matrix $A$ becomes available. The prediction of tumor motion using selected markers only requires a simple matrix
multiplication and hence the prediction can be achieved in a negligible amount of computation time. From Table \ref{Table:Validation:Real}, it is also found that the computational
time for marker selection varies from case to case, which is possibly ascribed to the different patient sizes.

\subsection{Comparison with ROI}
\begin{figure}[htp]
\centering
     \includegraphics[height=3.8in, width=3.8in]{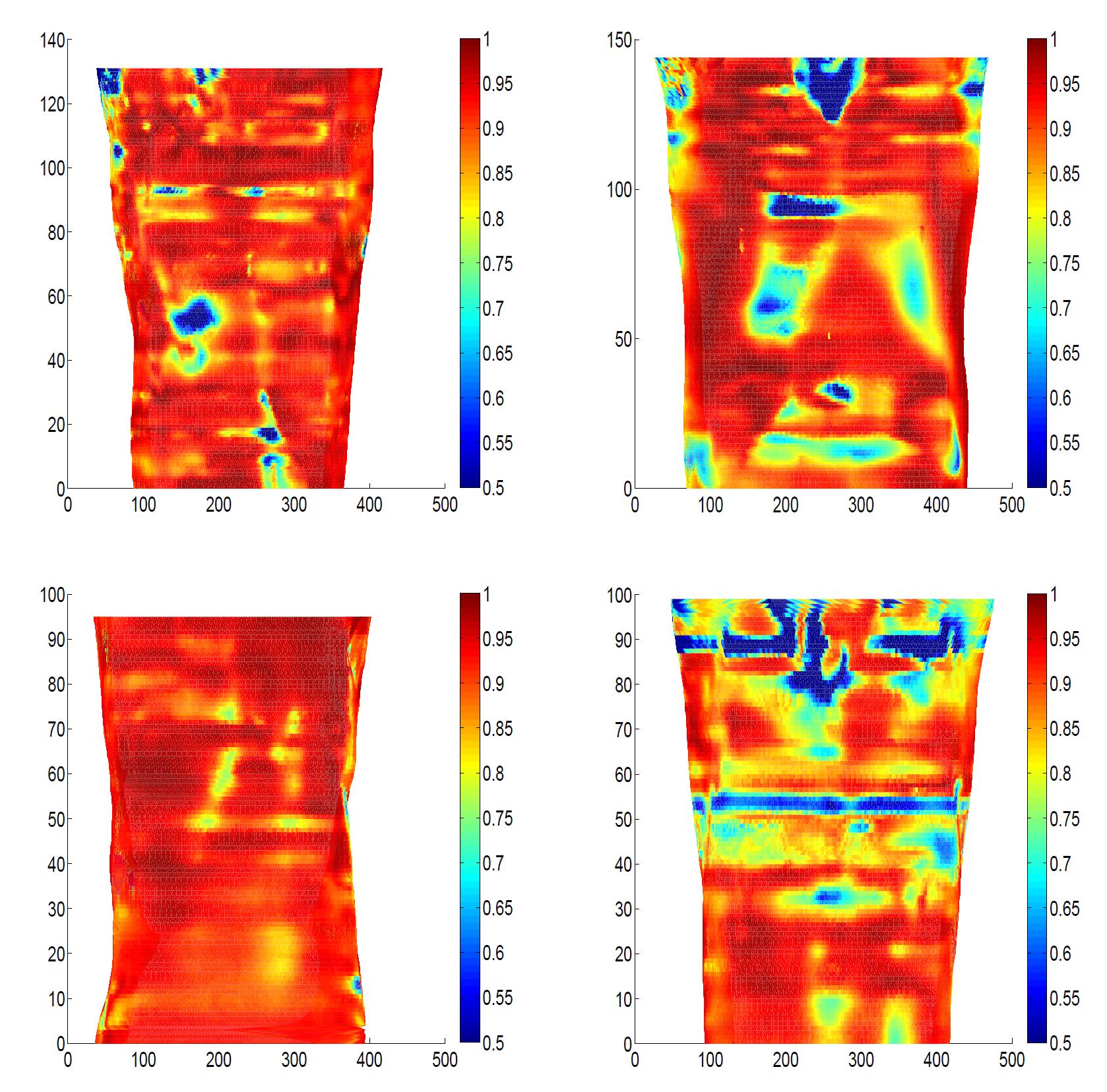}
    \caption{Color maps showing the correlation coefficients between the external surface motion and the internal tumor motion for 4 patients.}\label{Fig:Validation:correlation}
\end{figure}

\begin{figure}[htp]
\centering
     \includegraphics[height=3.8in, width=3.8in]{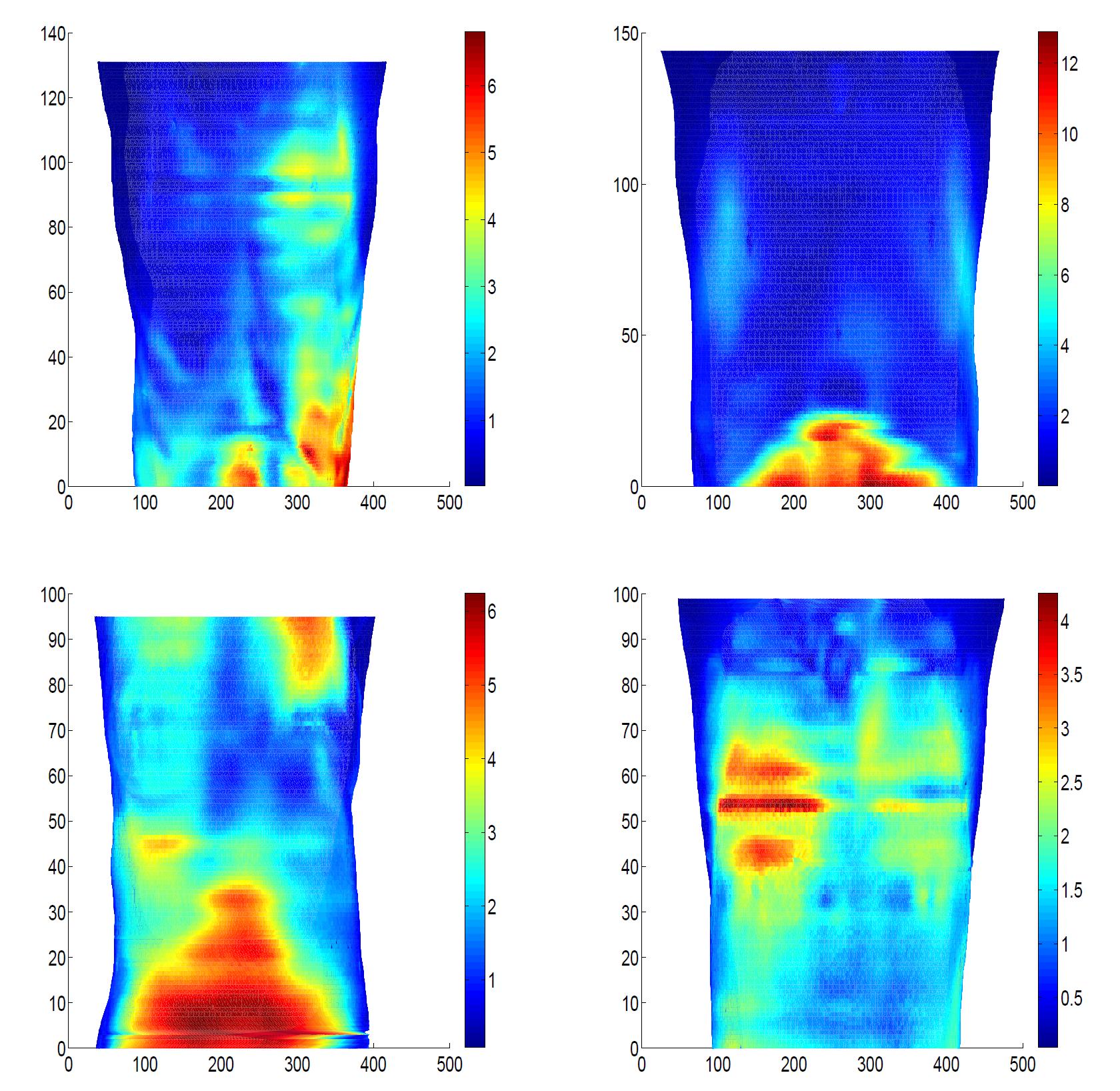}
    \caption{Color maps showing the amplitude of external surface motion for 4 patients.}\label{Fig:Validation:amplitude}
\end{figure}

\begin{figure}[htp]
\centering
     \includegraphics[height=3.8in, width=3.8in]{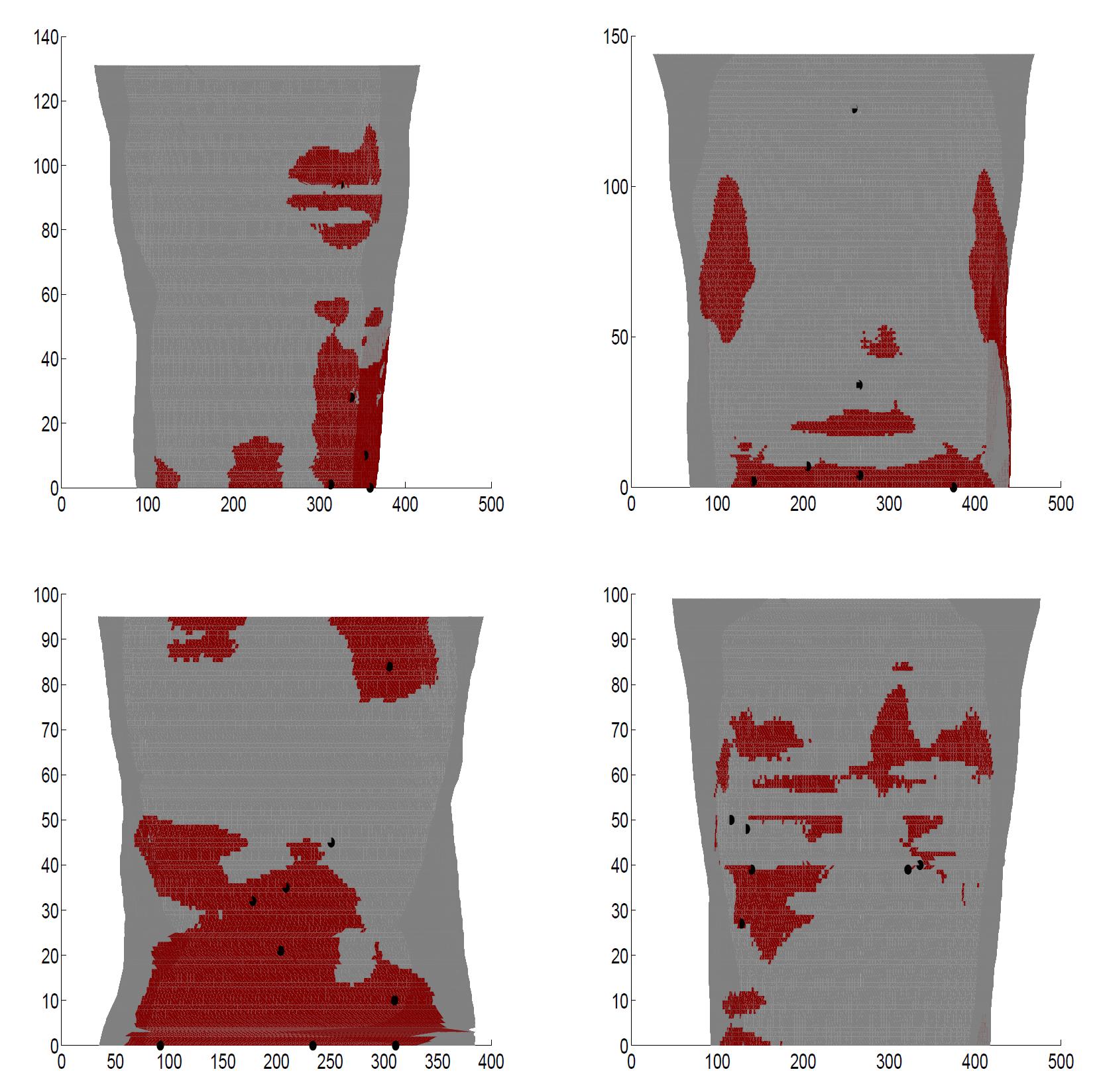}
    \caption{Color maps showing the regions of interest (where the motion amplitudes are relatively large and the correlation coefficients are relatively high) and the locations of
    the selected markers.}\label{Fig:Validation:marker}
\end{figure}

The correlation between the internal tumor motion in the S-I direction and the external surface motion is shown on Fig.~\ref{Fig:Validation:correlation}. In
Fig.~\ref{Fig:Validation:amplitude}, we also present the amplitude of external surface motion. Combining the correlation map and the motion amplitude map, the ROIs for each patient
can be identified, shown as red regions in Fig.~\ref{Fig:Validation:marker}, where the ROIs have correlation coefficients larger than 0.85  and surface motion amplitude greater than
$80\%$ of the maximum value. Apparently, the ROIs are highly dependent on different breathing motion patterns among patients. We also plot in
Fig.~\ref{Fig:Validation:marker} the locations of markers selected with our algorithm. We can see that most of the marker positions selected by our algorithm fall inside or close to
the
ROIs, which indicate the robustness of our algorithm.

\section{Conclusions}

In this paper, we proposed a novel mathematical model to automatically determine the optimal number and locations of
fiducial markers on patient's surface for predicting lung tumor motion. We also introduced an efficient numerical algorithm for solving the proposed model.  Experiments on the 4DCT
data of 4 lung cancer patients have shown that, by using our method, usually 6-7 markers are selected on patient's external surface. Most of these markers are in the regions where the
surface motion is relatively large and the correction between the surface motion and the internal tumor motion is relatively high. Using these markers, the lung tumor positions can be
predicted with an average 3D error of approximately $1mm$. Both the number of markers and the prediction accuracy are clinically acceptable, indicating that our method can be used in
clinical practice.

\section*{Acknowledgements}
This work is supported in part by the Master Research Agreement from Varian Medical Systems, Inc..

\section*{References}

\bibliographystyle{jmr}
\bibliography{ReferenceLibrary}

\end{document}